\documentclass[iop]{emulateapj}
\usepackage{graphics}
\usepackage{epsfig}
\usepackage{amsmath}

\newcommand{\degree}{\hbox{$^\circ$}}

\newcommand{\ltsimeq}{\la}
\newcommand{\gtsimeq}{\ga}
\newcommand{\msun}{M$_{\odot}$}

\newcommand{\HI}{H{\sc i}}

\renewcommand*{\thefootnote}{\fnsymbol{footnote}} 

\shortauthors{McQuinn, Mitchell, \& Skillman}
\shorttitle{The Panchromatic STARBurst IRregular Dwarf Survey (STARBIRDS) Data}

\begin{document}
\title{The Panchromatic STARBurst IRregular Dwarf Survey (STARBIRDS): Observations and Data Archive\protect\footnotemark[*]}
\footnotetext[*]{Based on observations made with the NASA/ESA Hubble Space Telescope, and obtained from the Hubble Legacy Archive, which is a collaboration between the Space Telescope Science Institute (STScI/NASA), the Space Telescope European Coordinating Facility (ST-ECF/ESA) and the Canadian Astronomy Data Centre (CADC/NRC/CSA).}
\author{Kristen B.~W.~McQuinn\altaffilmark{1}, 
Noah P.~Mitchell\altaffilmark{1,2,3},
Evan D.~Skillman\altaffilmark{1}
}

\altaffiltext{1}{Institute for Astrophysics, University of Minnesota, 116 Church Street, S.E., Minneapolis, MN 55455,\ {\it kmcquinn@astro.umn.edu}} 
\altaffiltext{2}{Department of Physics, St.~Olaf College, 1520 St. Olaf Ave, Northfield, MN 55057}
\altaffiltext{3}{Current address: Department of Physics and the James Frank Institute, University of Chicago, 929 East 57th Street, Chicago, Illinois 60637}

\begin{abstract}
Understanding star formation in resolved low mass systems requires the integration of information obtained from observations at different wavelengths. We have combined new and archival multi-wavelength observations on a set of 20 nearby starburst and post-starburst dwarf galaxies to create a data archive of calibrated, homogeneously reduced images. Named the panchromatic ``STARBurst IRregular Dwarf Survey'' (STARBIRDS) archive, the data are publicly accessible through the Mikulski Archive for Space Telescopes (MAST). This first release of the archive includes images from the Galaxy Evolution Explorer Telescope (GALEX), the Hubble Space Telescope (HST), and the \textit{Spitzer Space Telescope (Spitzer)} MIPS instrument. The datasets include flux calibrated, background subtracted images, that are registered to the same world coordinate system. Additionally, a set of images are available which are all cropped to match the HST field of view. The GALEX and $Spitzer$ images are available with foreground and background contamination masked. Larger GALEX images extending to 4 times the optical extent of the galaxies are also available. Finally, HST images convolved with a 5\arcsec\ point spread function and rebinned to the larger pixel scale of the GALEX and $Spitzer$ 24 $\micron$ images are provided. Future additions are planned that will include data at other wavelengths such as $Spitzer$ IRAC, ground based H$\alpha$, $Chandra$ X-ray, and Green Bank Telescope \HI\ imaging.
\end{abstract} 

\keywords{galaxies: dwarf --- galaxies: evolution --- galaxies: individual (Antlia dwarf, ESO~154-023, UGC~4483, UGC~6456, UGC~9128, NGC~625, NGC~784, NGC~1569, NGC~2366, NGC~4068, NGC~4163, NGC~4214, NGC~4449, NGC~5253, NGC~6789, NGC~6822, IC~2574, IC~4662, DDO~165, Holmberg~II) --- galaxies: starburst}

\section{Introduction \label{intro}}
The term ``starburst'' refers to a finite period of unsustainable star formation in a galaxy \citep[e.g.,][]{Gallagher1993, Heckman1997, Kennicutt2005}. The starburst phenomenon has been measured to last $>100$ Myr in nearby dwarf galaxies \citep{McQuinn2010b}, persisting much longer than previously thought based on measurements of individual, ``flickering'' pockets of star formation \citep{Schaerer1999, Mas-Hesse1999, Thornley2000, Tremonti2001}. The spatial distribution of the star formation ranges from being centrally concentrated to distributed across the optical disk of a galaxy, supporting the paradigm of a longer-lived, more global event \citep{McQuinn2012}. 

The two aforementioned studies measuring the temporal and spatial characteristics of starbursts \citep{McQuinn2010b, McQuinn2012} are based on reconstructing the star formation histories of dwarf galaxies from optically resolved stellar populations. Thus, the classification of a galaxy as a starburst does not rely on absolute star formation rate (SFR) thresholds, but, instead, considers the star formation properties in the context of a host galaxy's own evolutionary history. In this instance, a lower mass galaxy may host a significant star forming event even though the starburst would not compare in absolute terms with a starburst event in a more massive galaxy. Additionally, these studies are able to distinguish between active and fossil bursts. This is an important distinction as it allows for the study of both currently bursting systems and post-burst galaxies. 

\begin{figure*}
\plotone{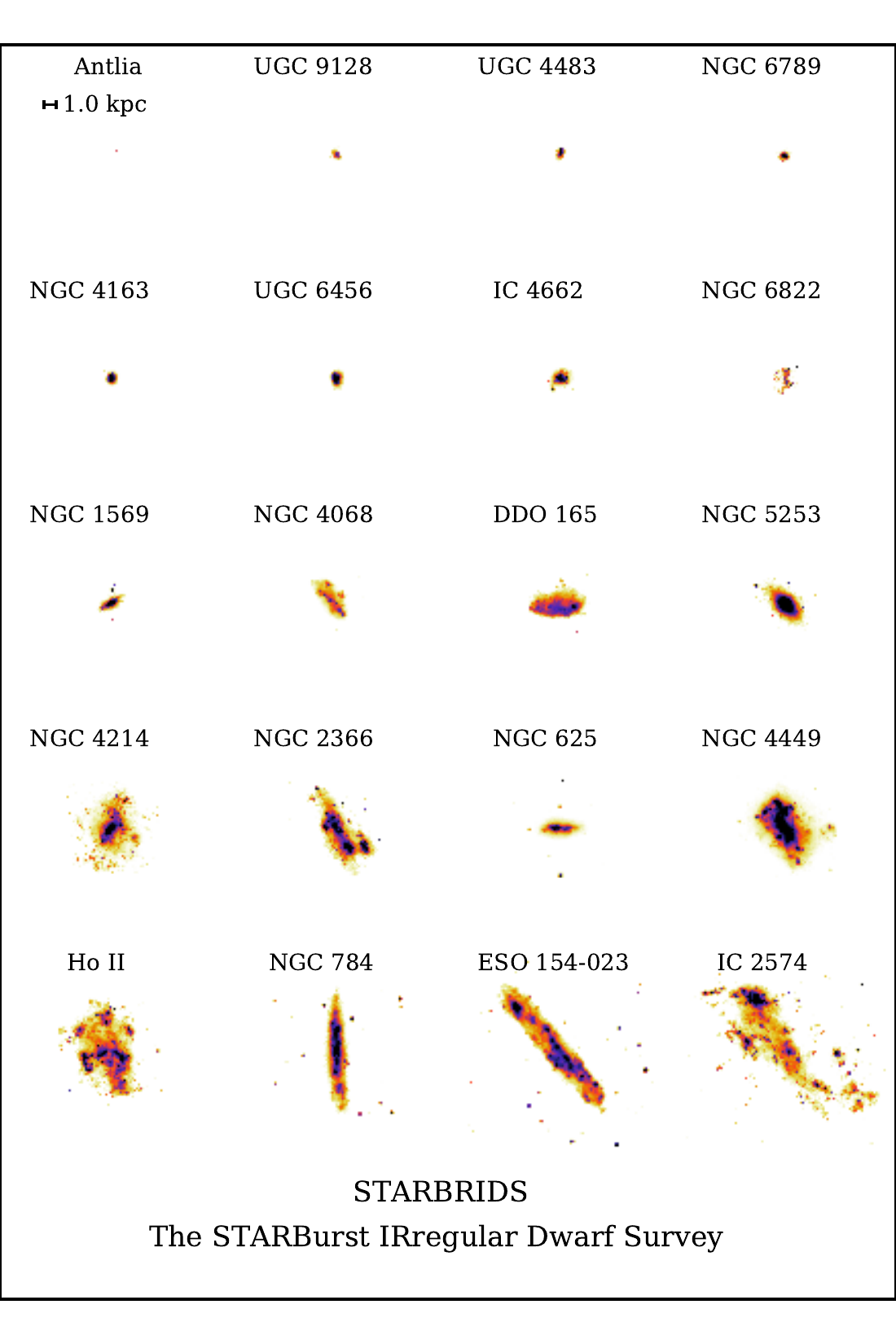}
\caption{GALEX NUV images of the STARBIRDS sample resampled to the same physical scale. The scale is noted in the upper left corner. The galaxies span nearly three decades in stellar mass and nearly four decades in SFR.}
\label{fig:grid}
\end{figure*}

As part of an on-going study of nearby starburst dwarf galaxies, we have processed new and archival multi-wavelength observations of a set of 20 starburst and post-starburst dwarf galaxies within 6 Mpc. As a natural result of this work, we have created the panchromatic ``STARBurst IRregular Dwarf Survey'' (STARBIRDS) data archive of processed images to facilitate further research with minimal image manipulation and processing. Figure~\ref{fig:grid} present near ultraviolet (NUV) images of the sample from the Galaxy Evolution Explorer (GALEX) telescope. The data were resampled to the same physical scale to show the range in galaxies probed by the survey.

This first public release of the data archive, hosted through the Mikulski Archive for Space Telescopes (MAST), includes far ultraviolet (FUV) and NUV imaging from the GALEX (Section~\ref{galex_data}), optical imaging from the Hubble Space Telescope (HST; Section~\ref{hst_data}), and infrared (IR) imaging from the $\textit{Spitzer Space Telescope}$ Multiband Imaging Photometer (MIPS; Section~\ref{mips_data}). The following data products are available$\colon$

\begin{itemize}
\item HST $V$ and $I$ band, GALEX FUV and NUV, and $Spitzer$ MIPS datasets that are cleaned, flux calibrated, matched to the same field of view, and registered to the same coordinate system. 
\item GALEX and $Spitzer$ images with foreground contamination and background galaxies fully masked. 
\item HST observations that are convolved to the GALEX 5\arcsec\ point spread function (PSF), rebinned to the 1.5\arcsec\ pixel scale of the GALEX and MIPS 24$\micron$ images, and registered to the same world coordinate system. These observations provide a direct comparison of morphological features at optical wavelengths to those observed at UV and mid-infrared wavelengths.
\item GALEX images with a larger field of view than the HST images extending to 4 times the diameter of the B-band 25 mag arcsec$^{-2}$ isophote in each system \citep{Karachentsev2004}. 
\end{itemize} 

Future additions to the archive are planned that will include $Chandra$ Space Telescope X-ray, $Spitzer$ IRAC, ground based H$\alpha$, and Green Bank Telescope \HI\ imaging. Additional datasets at complementary wavelengths (e.g., radio continuum observations) are also envisioned. 

The paper is organized as follows. Section~2 describes the galaxy sample, Section~3 describes the GALEX (FUV and NUV bands), HST ($V$ and $I$ band), and $Spitzer$ (MIPS bands) observations, and Section~4 describes the data processing at each wavelength. Section~5 lists the fields of view chosen for the archive, Section~6 explains the removal of foreground and background objects in the images and the flux measurements. Section~7 lists the additional data products available including HST images convolved to a larger PSF. Finally, Section~8 provides a summary of the work. Analysis of the GALEX, HST, and $Spitzer$ imaging is presented in a companion paper \citep{McQuinn2015}.

\section{The Sample \label{sample}}
\subsection{The STARBIRDS Galaxies}
Table~\ref{tab:galaxies} lists the dwarf irregular galaxies that comprise the STARBIRDS sample and their basic properties. The sample includes known starburst galaxies, post-starburst galaxies, and galaxies newly identified as starbursts based on the significant populations of young massive stars found in their optical color magnitude diagrams (CMDs) \citep{McQuinn2010a}. The sample was originally selected based on observations available in the HST archive that were suitable for measuring the duration of starbursts in dwarf galaxies \citep{McQuinn2009, McQuinn2010a}. With this goal in mind, optical imaging was chosen to meet three conditions. First, both $V$ and $I$ band images were required for each system. Second, the $I$ band observations were required to reach a minimum photometric depth of $\sim$2 mag below the tip of the red giant branch; a requirement for accurately constraining the recent star formation history of a galaxy \citep{Dolphin2002, McQuinn2010b}. Third, the galaxies were close enough such that HST imaging instruments are able to resolve their stellar populations (D$\ltsimeq$ 6 Mpc). These criteria provided an appropriate dataset in which to study the temporal and spatial characteristics of starbursts. 

The strength of this HST archival-defined sample is that the dwarf galaxies cover a range in luminosity ($-17.94<$ M$_B < -9.80$), morphology (SBdm, IAB(s)m, pec, IBm), and metallicity ($7.39 \leq 12 +$(O/H)$ \leq 8.38$). The majority of the galaxies have high Galactic latitudes and low foreground extinction \citep[i.e., A$_R \ltsimeq 0.2$;][]{Schlegel1998}. NGC~1569 and NGC~6822 are two exceptions, with $A_R$ extinction values of 1.9 and 0.6 mag, respectively. Note, however, that the sample is not a homogeneous, volume limited sample. Thus, care must be taken when interpreting the results from the sample.

\subsection{General Properties of the Sample}
\begin{figure}
\includegraphics[width=0.49\textwidth]{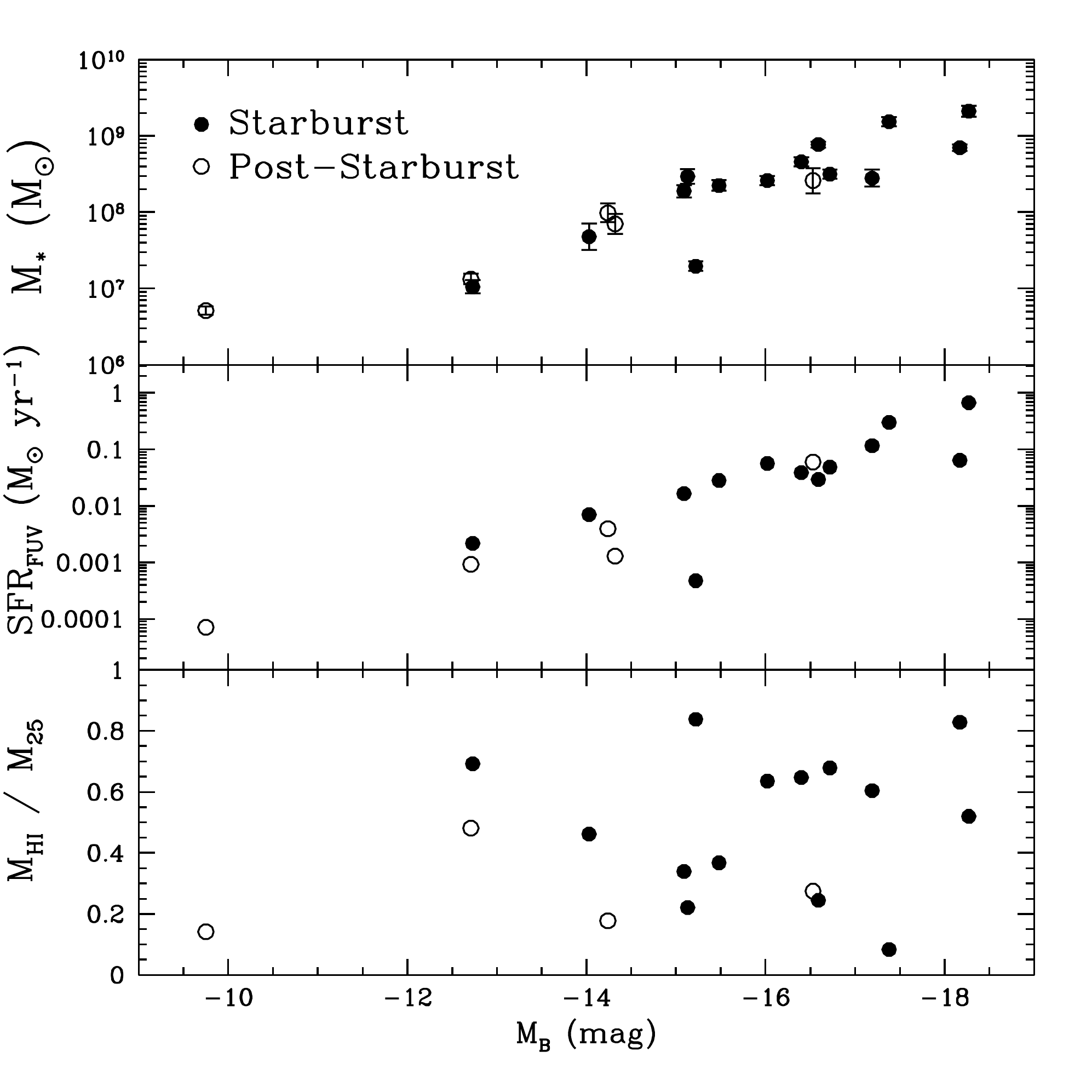}
\caption{Stellar mass, FUV SFRs, and fraction of neutral hydrogen to ``total'' mass \citep[i.e., M$_{HI}$/M$_{25}$;][]{Karachentsev1999}  of the sample as a function of absolute B-band magnitudes. Starburst galaxies are plotted as filled circles; post-starburst galaxies are plotted as unfilled circles (see Table~\ref{tab:galaxies}). No FUV data are available for IC~4662 (see Section~\ref{galex_data} 3.1); M$_{HI}$/M$_{25}$ was not available for NGC~6789. While the sample spans a range in luminosity, metallicity, gas mass fraction, and SFR, it was originally designed from an archival study and is not a volume limited sample. There is a clear trend between stellar mass and luminosity;  the outliers are NGC~6822, Ho~II, and NGC~4214, whose stellar disks are only partially covered in the HST observations from which the stellar masses were derived.}
\label{fig:mb}
\end{figure}

Figure~\ref{fig:mb} presents three measured properties as a function of the absolute B-band magnitudes of the sample, including the stellar mass of the galaxies in the $HST$ field of view (top panel), the SFR calculated from the GALEX FUV emission (middle panel), and the fraction of neutral hydrogen to total mass within the Holmberg radius \citep[bottom panel;][]{Karachentsev2004}. Starburst galaxies are plotted with filled circles, while post-starburst galaxies are plotted with unfilled circles. 

The stellar mass estimates in the top panel of Figure~\ref{fig:mb} were derived from the lifetime-average SFRs from \citet{McQuinn2010b} with a 30\% correction for the amount of mass returned to a galaxy over the lifetime of the stellar populations \citep{Kennicutt1994}. As expected, there is a general trend between stellar mass and luminosity. Note, however, that while the majority of the galaxies have a small enough angular size to have their stellar mass well-represented within in the HST field of view, the optical disk of three galaxies significantly extend beyond the observational footprints (see Table~1). These three galaxies, NGC~6822 ($M_B = -15.22$), Ho~II ($M_B = -16.72$), and NGC~4214 ($M_B = -17.19$), show deviations from the general trend as the stellar mass is derived from a region that is smaller the region encompassed by the luminosity measurements. 

The SFRs in the middle panel of Figure~\ref{fig:mb} were calculated from the extinction corrected FUV fluxes using the scaling relation from \citet{Hao2011}. Because the FUV emission originates primarily from the photospheres of young stars of intermediate and high mass (M$\gtsimeq3$ \msun), it is a direct tracer of the recent star formation (t $\sim100$ Myr). The most widely used UV scaling relation, presented in \citet{Kennicutt1998}, is based on the theoretical results from \citet{Madau1998} over the wavelength range 1500-2800 \AA. Recently, the theoretical relation was re-calibrated using updated stellar evolution libraries and spectral models \citep{Hao2011, Murphy2011} with the Starburst99 code \citep{Leitherer1999, Vazquez2005}. The resulting revision is $\sim5$\% lower than the original value from \citet{Kennicutt1998} for the same IMF. Uncertainties on the \citet{Hao2011} FUV-SFR relation are not well-quantified. Note that the theoretical scaling relation applied here may under-estimate the recent SFRs in these systems by 65\% or more. We present a full comparison of the FUV-based scaling relation with CMD-based SFRs, including quantified uncertainties on the FUV-SFR calibration in our companion paper \citep{McQuinn2015}. No FUV data are available for one galaxy, IC~4662 (see Section 3.1). Similar to the top panel, there is a general trend between SFR and luminosity in the sample. 

The fraction of neutral hydrogen to ``total'' mass in the bottom panel of Figure~\ref{fig:mb} is from \citet{Karachentsev1999}. In that study, the hydrogen mass was calculated based on \HI\ flux measurements; the total mass inside the Holmberg radius was estimated based on the galaxy rotation amplitude, corrected for inclination and turbulence. There is no apparent trend between the fraction of hydrogen to total mass and overall galactic luminosity.

\begin{figure}
\includegraphics[width=0.49\textwidth]{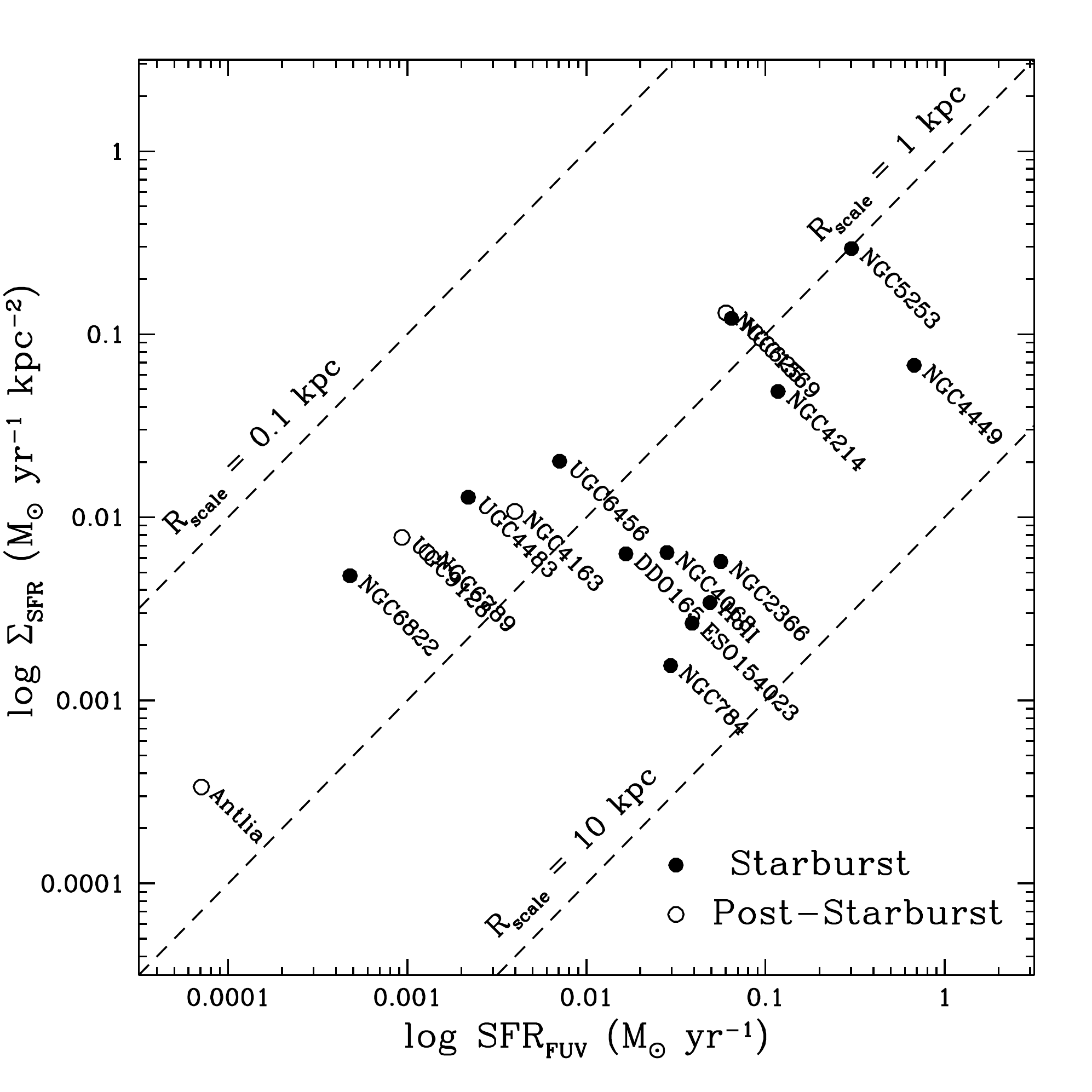}
\caption{Distribution of FUV SFR densities and FUV SFRs for the sample. Starburst galaxies are plotted as filled circles; post-starburst galaxies are plotted as unfilled circles. Dashed lines denote star forming regions of constant radii (R$_{scale}$) from 0.1 to 10 kpc.}
\label{fig:ssfr}
\end{figure}

Figure~\ref{fig:ssfr} compares the SFR density based on SFRs derived from the GALEX FUV emission and the area containing the majority of the FUV emission, with the SFRs derived from the GALEX FUV emission using the calibration from \citet{Hao2011}. The dashed lines denote star forming regions of constant radius from 0.1 to 10 kpc from top to bottom. Starburst galaxies are represented as filled circles; post-starburst galaxies are unfilled circles. The SFR density spans a wide range, reflecting the diversity of the star formation properties in the sample. All galaxies show elevated levels of recent or ongoing star formation levels based on their historical averages, thus representing the low-mass end of the starburst galaxies.

\begin{figure}
\includegraphics[width=0.49\textwidth]{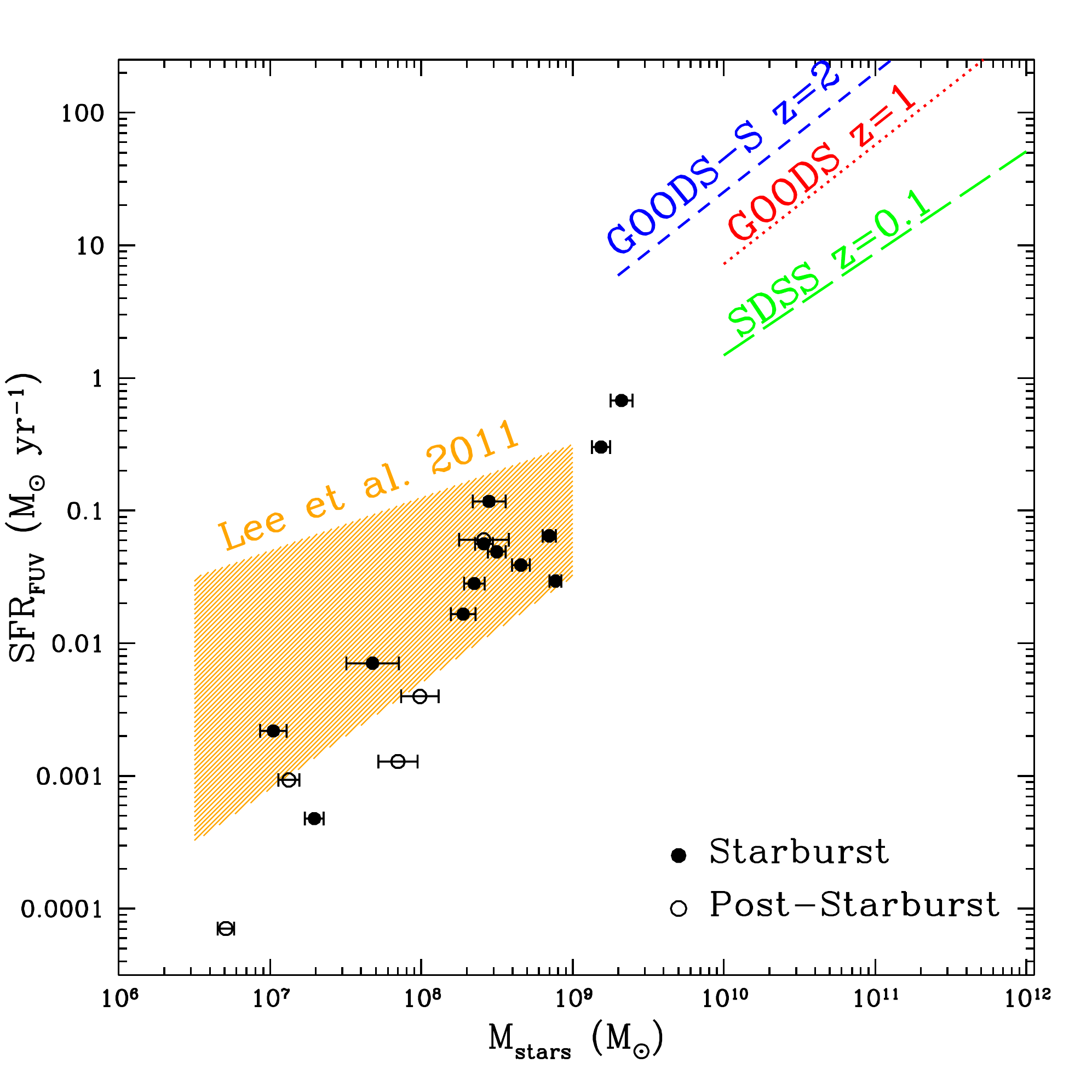}
\caption{Comparison of the FUV SFRs with the stellar mass for the sample. Starburst galaxies are plotted as filled circles. Post-starburst galaxies, plotted as unfilled circles, predominate at the lower SFRs. Our sample probes the low mass galaxy regime and overlaps with the sample biased towards dwarf galaxies with higher recent SFRs studied by \citet{Lee2011}. The majority of dwarf Irregular galaxies in their survey lie in the region shaded in orange. Best fit lines from surveys of more massive galaxies are also shown including GOODS-S \citep[$z = 1$; short dashed blue line][]{Daddi2007}, GOODS \citep[$z = 1$; dotted red line][]{Elbaz2007}, and SDSS \citep[$z = 0.1$; long dashed green line][]{Brinchmann2004} surveys. The range in SFR and stellar mass in our sample highlights the diversity of the galaxies which include both active and post-starbursts.}
\label{fig:mass_sfr}
\end{figure}

Figure~\ref{fig:mass_sfr} compares the SFRs derived from the GALEX FUV emission with the stellar masses of the galaxies. Post-starburst galaxies are shown as unfilled circles and predominate at the lower SFRs. A sample of dwarf Irregular galaxies spanning an overlapping range in FUV based SFRs and stellar mass range, highlighted in shaded orange, was studied by \citet{Lee2011}. Their sample was based on a sensitivity limit which probes higher SFRs in these lower mass galaxies. Also shown are best fit lines from a number of other surveys with samples of more massive galaxies at $z=0$ from the SDSS \citep[long dashed green line;][]{Brinchmann2004}, at $z=1$ from the GOODS survey \citep[dotted red line;][]{Elbaz2007}, and at $z=2$ from the GOODS-S \citep[short dashed blue line;][]{Daddi2007}. Note that the axes of the plot were chosen to cover the range in SFR and stellar mass in our data, as well as the higher rates of star formation in the more massive systems of the high redshift surveys.

\section{The Observations \label{observe}}
\subsection{GALEX UV Imaging \label{galex_data}}
The GALEX instrument provides simultaneous far ultraviolet (FUV; 1350-1750~\AA) and near ultraviolet (NUV; 1750-2750~\AA) imaging using a dichroic beam splitter. The native pixel resolution for each detector is 1.5\arcsec\ pixel$^{-1}$. The GALEX instrument is described in detail in \citet{Martin2005}, and its in-orbit performance and calibration by \citet{Morrissey2005, Morrissey2007}. 

New, deep GALEX observations (PID 60026) were obtained with the goal of reaching a uniform exposure time of 10.5 ksec (i.e., 7 orbits) in each of the GALEX UV bands when combined with existing archival data. This integration time is equivalent to reaching an FUV photometric depth of 22.5 mag, corresponding to the surface brightness expected from a constant SFR per unit area of 8$\times10^{-4}$ \msun~yr$^{-1}$ kpc$^{-2}$ at a distance of 3 Mpc. This depth was targeted to facilitate an exploration of the outer regions in starburst galaxies where lower SFRs predominate. Preceding the execution of the new observations, the FUV detector aboard GALEX failed. Thus, the data for the FUV band is comprised solely of archival data with integration times ranging from $\sim1.5$ ksec (1 orbit) to $\sim21$ ksec (14 orbits). One system, IC~ 4662, did not have archival observations and therefore lacks FUV observations; the NUV observations were executed as planned. We include IC~4662 NUV data for completeness. Further, due to a calibration change that occurred as a result of Coarse Sun Point event on May 4th, 2010\footnote{http://www.galex.caltech.edu/wiki/Public:Documentation}, some of the affected observations failed the quality assurance tests and are not included. Table~\ref{tab:uv_observations} lists the final GALEX exposure time achieved per system.

\subsection{HST Optical Imaging \label{hst_data}}
This study used observations from two instruments aboard the HST, the Advanced Camera for Survey instrument \citep[ACS,][]{Ford1998} and the Wide Field Planetary Camera 2 instrument \citep[WFPC2,][]{Holtzman1995}. Details on the HST observations, including exposure times are presented in Table~\ref{tab:hst_observations}. The ACS observations were taken with the wide-field camera's $V$ filter (F555W or F606W) and $I$ filter (F814W). The ACS instrument has a native pixel scale of 0.05\arcsec\ pixel$^{-1}$ and a field of view covering 202\arcmin$\times202$\arcmin. The WFPC2 observations were taken with the $V$ filter (F555W) and the $I$ filter (F814W). The WFPC2 instrument has three wide field CCDs, each providing an 800$\times$800 pixel format with a pixel scale of 0.1\arcsec\ pixel$^{-1}$, and one planetary camera CCD providing the same pixel format but with a smaller pixel scale of 0.05\arcsec\ pixel$^{-1}$ and a correspondingly smaller field of view. 

\subsection{$Spitzer$ IR Imaging \label{mips_data}}
We used archival images from the $Spitzer$ obtained in the mid-infrared regime from the MIPS instrument \citep{Rieke2004} at 24, 70, and 160 $\micron$. The MIPS instrument produces images with fields of view for the 24, 70, and 160 $\micron$ bands of 5\arcmin$\times$5\arcmin, 2.5\arcmin$\times$5\arcmin, and 0.5\arcmin$\times$5\arcmin\ respectively, resampled and mosaicked with corresponding pixel scales of 1.5, 4.5, and 9.0\arcsec\ pixel$^{-1}$. 

Table~\ref{tab:mips_observations} lists the program ID of the MIPS observations and the total exposure times for each galaxy. Sixteen of the twenty galaxies in the sample were originally observed as part of either the Local Volume Legacy Survey (LVL) \citep{Dale2009} or the Spitzer Infrared Nearby Galaxy Survey (SINGS) \citep{Kennicutt2003, Regan2004} projects. These MIPS observations were taken using a scan-mapping mode in two separate visits allowing for the removal of detector artifacts and transient objects (e.g., asteroids) in the final images. Observing in the scan-map mode causes integration times to vary per pixel in each image. The exact observational depth is provided in a weight map for each galaxy with an average exposure time in the center pixels listed in the image headers. For the remaining four galaxies, the MIPS observations available from the Spitzer Heritage Archive (SHA) were incomplete. The observations were short (t$<10$ sec) and, in the case of IC~4662, did not cover the HST field of view. In addition, the archival data were available at only one or two of the three MIPS bandpasses. 

\section{Data Processing\label{data_process}}
\subsection{GALEX Data Processing}
The FUV and NUV data were processed through the GALEX pipeline (v7.1) which converts the GALEX satellite telemetry data (i.e., photon lists) into calibrated intensity maps in counts per second. The v7.1 pipeline accounts for the change in the NUV calibration due to the Coarse Sun Point event on May 4th, 2010. The intensity maps include corrections from flat-fielding and account for a drift in the photometric calibration (0.25\% and 1.5\% fainter per year for the FUV and NUV bands, respectively). The GALEX pipeline also co-adds the archival and new observations into one FUV and one NUV image per target. As the original observations used to create the deeper, co-added tiles can have slightly varying fields of view, the GALEX pipeline provides high resolution relative response maps which measure the depth of the observations per pixel. Table~\ref{tab:uv_observations} provides the tile names for the co-added intensity maps.

In addition to a co-added intensity map, the GALEX data pipeline also provided a background image and a background subtracted intensity map \citep[in counts s$^{-1}$;][see \S3.3]{Morrissey2007}. For most observations, the pipeline background images provide a robust measurement of the sky flux and thus, the pipeline background subtracted images were used in our analysis and in the archive. However, for higher surface brightness galaxies, the background images included clearly visible structure from the galaxies. In addition, observations for one galaxy (Antlia Dwarf) showed cirrus contamination in the subtracted image. For these systems, we replaced the background flux in a $3\times$D$_{25}$ region centered on the target galaxies with the median flux value from a region of at least 1000 pixels within a $3\times$D$_{25} - 4\times$D$_{25}$ mag arcsec$^{-2}$ annulus of the galaxy, where $D_{25}$ is the major diameter measured to B-band 25 mag arcsec$^{-2}$ isophote reported in \citet{Karachentsev2004}. The galaxies for which we performed this custom background subtraction process are identified in Table~\ref{tab:uv_observations}. 

\subsection{HST Data Processing}
The HST data include single and multiple pointing observations. For all single pointing observations, we used the co-added images available through the Hubble Legacy Archive (HLA). For the five galaxies with multiple pointings (NGC~4449, NGC~5253, NGC~2366, IC~2574, and Ho~II), co-added mosaicked images of only one system (NGC~4449) were available through the HLA. For the remaining four galaxies, we created the co-added mosaicked images using MultiDrizzle software on the STScI python 2.7 platform. No additional processing was performed on the HST FITS mosaic images. These HST images provide the reference field of view of the star formation and are used for comparison at other wavelengths \citep{McQuinn2015}. Note that photometry is best performed on the raw images available through MAST, which are processed using the standard HST pipeline.

\subsection{MIPS Data Processing}
For the sixteen galaxies observed as part of the LVL or SINGS $Spitzer$ legacy programs, we used the enhanced data products provided by these programs. A detailed description of the data processing can be found in \citet{Dale2009} and in the data delivery documents of the surveys. Here, we list the main data reduction steps used by the LVL and SINGS team. 

The MIPS observations were processed using the MIPS Data Analysis Tool \citep[MIPS DAT][]{Gordon2005} and calibrated in MJy sr$^{-1}$. Data processing for the MIPS 24$\micron$ data included a droop correction, a non-linearity correction in the ramps, and a dark current correction. The images were flat-fielded, latent images were masked, and jailbar patterns corrected. The background was estimated using a fit with a third order polynomial and subtracted from the images. Scattered light was also subtracted. 

Data processing for the MIPS 70 and 160$\micron$ data included a linear fit to the ramps, which also removes cosmic rays and readout jumps, and applies an electronic non-linearity correction. Stimflash frames, or on-board calibration stimulator frames used to track the detector response carefully, were used for responsivity corrections. The images had dark current subtractions, illumination corrections, and drift corrections. Short-term variations caused by drift were subtracted and cosmic rays were masked. 

For the four galaxies that were not observed as part of these larger $Spitzer$ legacy programs, we created the MIPS mosaics from the data available through the SHA. Using the $Spitzer$ MOPEX software, we deleted the first frames (which were taken in the high dynamic mode), flat-fielded the observations, and created mosaic images from the archival data. The median background was estimated in a region outside of the galaxy using the IRAF task \textsc{imstat} and subtracted from the mosaic. The 70 $\micron$ observations for NGC~1569 and NGC~6789 were unusable due to the presence of stimflash in all frames. 

\section{Rectification and Cropping\label{fov}}
\begin{figure}
\includegraphics[width=0.45\textwidth]{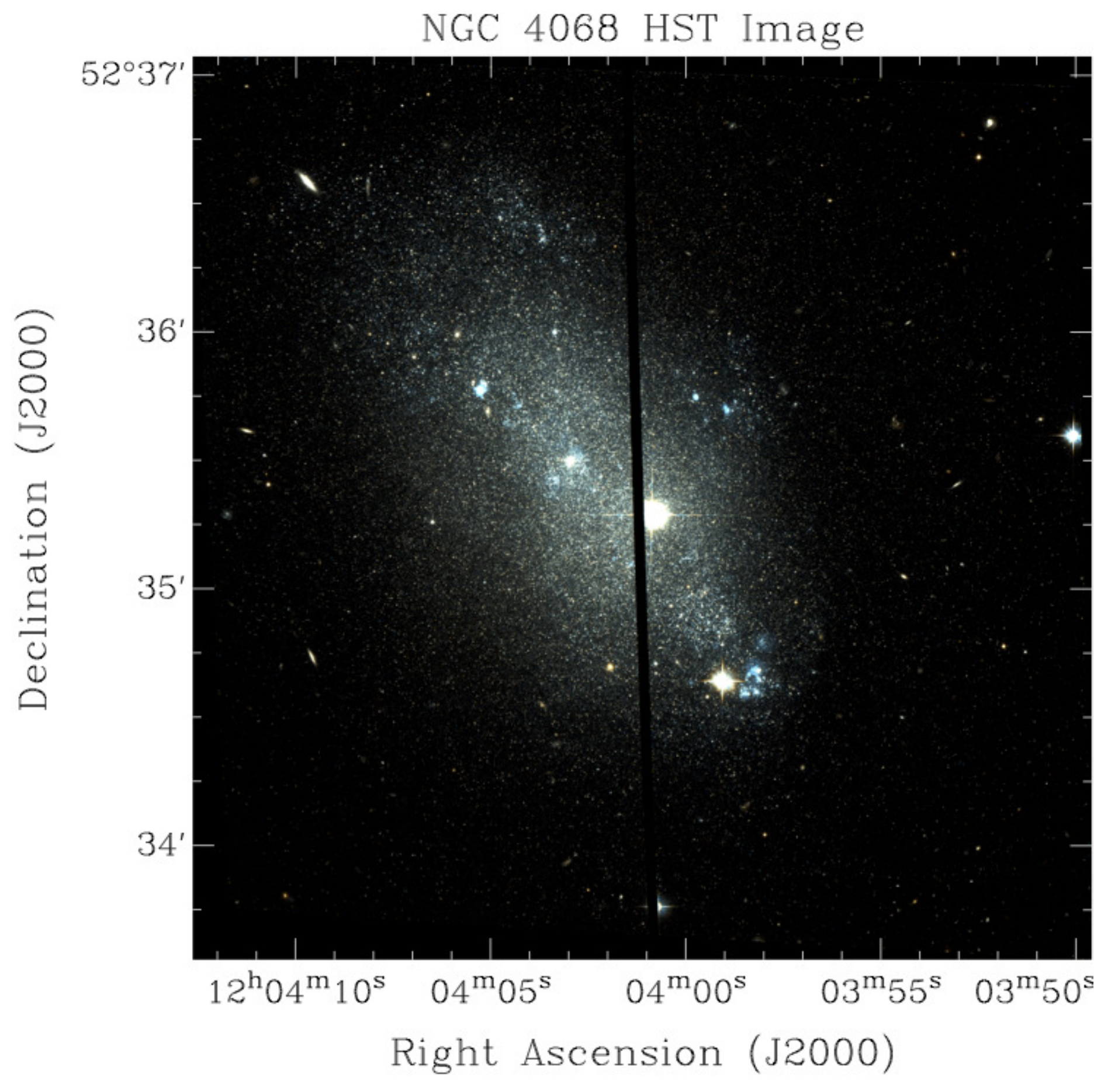}
\caption{A representative composite image from HST observations of a STARBIRDS galaxy, NGC~4068, created using the F606W image (blue), F814W image (red), and an averaged F606W and F814W image (green). North is up and East is left. The HST footprint defines the field of view selected for study in the GALEX and Spitzer MIPS images shown in Figures~\ref{fig:uv_image} and \ref{fig:mips_image}.}
\label{fig:hst_image}
\end{figure}

Because the data populating the archive were used in a comparison study of star formation at optical and UV wavelengths and potential extinction traced at infrared wavelengths, it was useful to crop the images to the same field of view. Not only do the HST images have the smallest field of view of the different instruments, the resolved stellar populations in the HST imaging also provide the basis for comparison with the star formation traced by the GALEX imaging. Thus, the HST images provided the reference field of view for the images at the other wavelengths. Figure~\ref{fig:hst_image} presents a representative image of the $HST$ data for the alaxy NGC~4068. However, one of the advantages of the GALEX detectors is the large 1.25\degree\ field of view. Therefore, a second field of view was chosen for the GALEX images that extends to fourfold the optical diameter of each galaxy.

\subsection{Matched HST Field of View}
\begin{figure}
\plotone{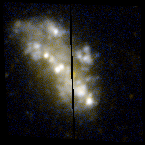}
\caption{A composite image of NGC~4068 from GALEX observations created using the FUV image (blue), the NUV image (red), and an averaged FUV and NUV image (green). The field of view matches that of the HST footprint shown in Figure~\ref{fig:hst_image}, including the chip gap of the HST ACS detector. }
\label{fig:uv_image}
\end{figure}

The first GALEX field of view covers the HST observational footprint. The GALEX images were cropped using the WCSTOOLS task \textsc{getfits} based on the center coordinates, full axes size, and orientation from the HST astrometry solution in the FITS headers. While these cropped, square images match the size of the HST FITS files, they do not match the actual HST fields of view as the ACS instrument is affected by distortion and the WFPC2 instrument has an irregularly shaped footprint. Thus, the cropped, square GALEX images were matched pixel-to-pixel to the HST observations; flux values in the UV images were set to zero if the corresponding locations in the HST images were not observed. For non-dithered, split orbit, HST ACS observations, the chip gap was also excluded in these cropped images. Note that the UV images are up to 1-pixel larger in length and width than the HST data due to the different pixel scales of the GALEX and HST data. Figure~\ref{fig:uv_image} presents a representative image of the GALEX data for the same galaxy shown in Figure~\ref{fig:hst_image}, NGC~4068. 

\begin{figure}
\includegraphics[width=0.49\textwidth]{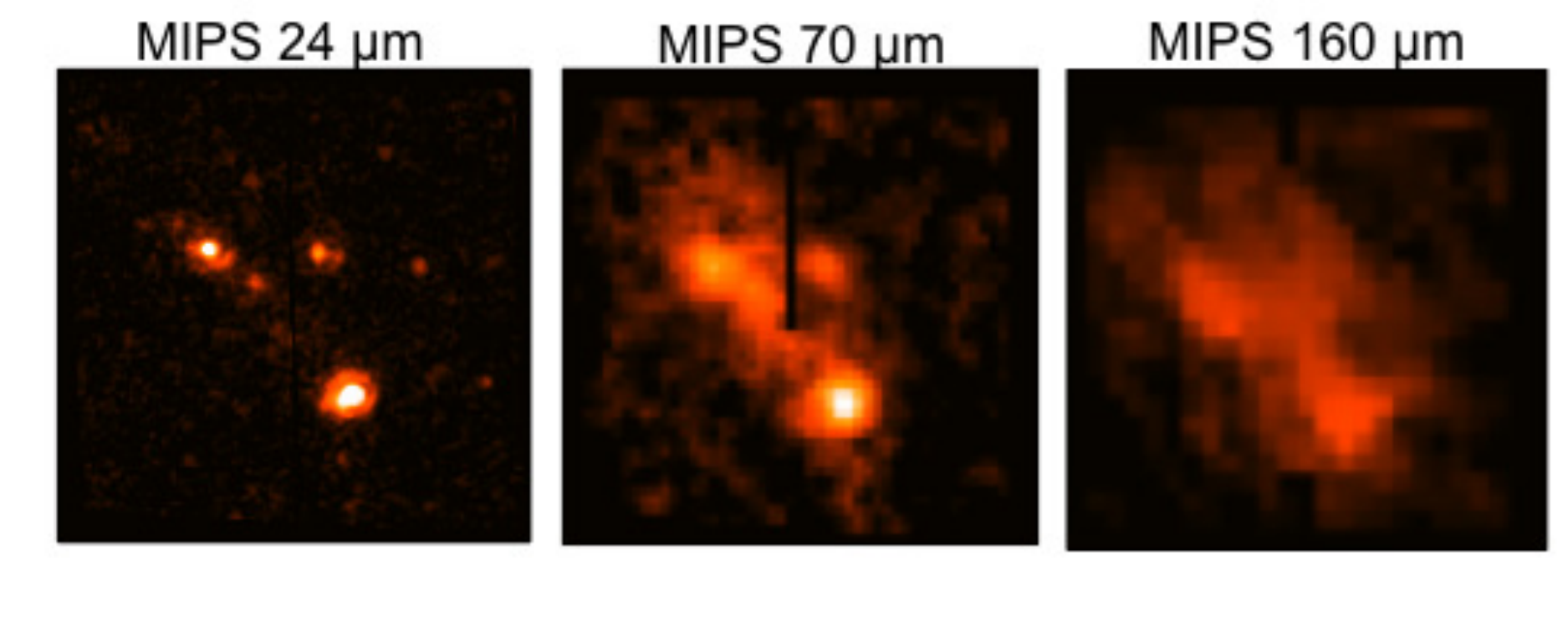}
\caption{Spitzer MIPS images of NGC~4068 at 24$\micron$, 70$\micron$, and 160$\micron$. The field of view matches that of the HST footprint shown in Figure~\ref{fig:hst_image}. Note how the 24$\micron$ image traces the star forming regions while the 160$\micron$ traces the more widespread diffuse emission from dust with the galaxy. }
\label{fig:mips_image}
\end{figure}

The MIPS images were cropped to the field of view of the HST observations following the same two step process employed for the GALEX data. First, a square cropped image was created using the WCSTOOLS task \textsc{getfits} based on the center coordinates, full axes sizes, and orientation from the HST astrometry solution in the FITS headers. Second, these cropped images were matched pixel-to-pixel to the HST observations; flux values in the MIPS images were set to zero if the corresponding location around the edges in the HST images or in the HST ACS chip gap were unobserved. Some of the MIPS fields of view did not encompass the entire HST field of view. In these cases, the MIPS total flux measurements are lower limits for the HST field of view. Figure~\ref{fig:mips_image} presents representative images of the  MIPS data for the same galaxy shown in Figures~\ref{fig:hst_image} and \ref{fig:uv_image}, NGC~4068.

\subsection{Full Galaxy GALEX Field of View}
The second GALEX field of view covers $4\times$D$_{25}$ for each galaxy. Whereas the field of view defined by the HST instruments does not cover the complete stellar disk of the more extended galaxies, this larger field of view provides complete areal coverage beyond the stellar disks on each system. The native 1.25\degree\ circular field of view of the GALEX images was cropped using the WCSTOOLS task \textsc{getfits} centered on the RA and Dec coordinates of each system with a length and width equal to $4\times$D$_{25}$. The coordinates and D$_{25}$ values are provided in Table~\ref{tab:galaxies}.

\section{Contamination Masking and Flux Measurements\label{flux}}
Background galaxies and foreground stars can affect the fidelity of the integrated flux measurements in the GALEX and MIPS images. Therefore, care was taken to mask these objects in each image cropped to the HST field of view in a multi-step process.

For the GALEX images, point sources with fluxes $\geq4~\sigma$ were identified in the NUV images using the IRAF task \textsc{starfind}. Second, these point sources were visually inspected in the FUV, NUV, and V-band images. Individual sources were flagged as contamination based on their morphology in the optical images (i.e., points sources that were identifiable in the optical images as background galaxies or foreground stars). For ambiguous objects, we performed a near-position search in the NASA/IPAC Extragalactic Database (NED) for cataloged stars and background galaxies. If the object in question was not a previously identified object, it was not masked. Third, when the masked point source did not overlap with an area of UV emission originating from the target galaxy, the flux values within a 3-pixel radius of the background or foreground object were replaced with the flux values from a neighboring region of pixels without UV emission using the IRAF task \textsc{imedit}. For more extended sources and for sources on the edge of the field of view, custom regions were used to ensure accurate replacement of all pixels affected by foreground or background contamination. Table~\ref{tab:uv_observations} lists the images that required masking.

We masked the background galaxies and foreground stars in the MIPS images using the same procedure used on the GALEX images. Briefly, point sources with fluxes $\geq4~\sigma$ were identified in the MIPS 24 $\micron$ images and cross-referenced with their optical counterparts in the V-band images. Contaminating point sources were identified based on their optical morphology. For ambiguous cases, we used the NED to help identify background galaxies by coordinates. The contaminating point sources were masked using flux measurements from neighboring sky pixels. The 24 $\micron$ list of point sources was also used to help identify sources in the 70 and 160 $\micron$ MIPS images. At these longer wavelengths, most background galaxies and foreground stars fall below the detection limit of the observations. Thus, only a few images required masking. In some of the 160 $\micron$ images, the lower spatial resolution made source identification ambiguous; no masking was performed in these instances. Table~\ref{tab:mips_observations} lists the images that required masking.

We measured the flux in the background subtracted, masked GALEX and MIPS images in the HST field of view using the IRAF\footnote{IRAF is distributed by the National Optical Astronomy Observatory, which is operated by the Association of Universities for Research in Astronomy (AURA) under a cooperative agreement with the National Science Foundation.} task \textsc{polyphot}. The task \textsc{polyphot} employs a user-defined polygon shaped aperture which matched the HST WFPC2 or ACS footprint, depending on the observations. The integrated UV flux measurements in the HST fields of view are listed in Table~\ref{tab:uv_observations}, the integrated MIPS flux measurements are listed in Table~\ref{tab:mips_observations}. Extinction corrected luminosities and SFRs based on these flux measurements are reported in our companion paper \citep{McQuinn2015}.

\section{Enhanced Data Products \label{enhanced_data}}
Additional sets of HST images were created at the resolution and pixel scale of the GALEX and $Spitzer$ MIPS 24 $\micron$ observations. The original HST images were convolved to a 5\arcsec\ PSF and re-binned to the a pixel scale of 1.5\arcsec, the native pixel scales of both the GALEX detectors and the MIPS 24$\micron$ detector. In addition, JPEG files were generated from all data FITS files available in the archive, including 2-color optical jpegs. Each enhanced data product is described below.

\subsection{Reprocessed \textit{HST} Images}
The HST $V$ and $I$ band images have a pixel scale of 0.05\arcsec\ for the ACS instrument and 0.1\arcsec\ for the WFPC2 instrument after processing by the Hubble Legacy Archive. The full width at half maximum (FWHM) of the PSF is $\sim2$ pixels for the HST images. Following the procedure employed by the NRAO Astronomical Image Processing System (AIPS) software, we convolved the smaller HST PSFs with the larger $\sim5$\arcsec\ PSF of the GALEX detectors. Once convolved with the larger PSF, we re-binned the images using the IDL task $\textsc{frebin}$ and mapped the images to same coordinate system using the world coordinate systems header information and the IDL tasks $\textsc{triangulate}$ and $\textsc{griddata}$. The resulting images enable a comparison of the optical morphological features with the morphological features at UV and MIR wavelengths. 

Comparing the reprocessed HST images with the GALEX images, there are numerous morphological features that are traced at both wavelengths, as expected. There are also notable differences between some of the features seen in one and not the other image. In contrast, while the reprocessed HST images and the MIPS 24$\micron$ images often trace the same star forming regions, the details of the morphological features are generally different. These similarities and differences are primarily due to the different sources of the emission at UV (massive, young stars), optical (stars of varying masses and ages) and MIR (dust heated by stars) wavelengths. Detailed analysis of the star formation at these 3 wavelength regimes are reported in a companion paper \citep{McQuinn2015}.

\subsection{JPEG Preview Images}
Grayscale JPEG files were created using the Montage software routine \textsc{mJPEG} for all of the fits files. The JPEG files were made using a Gaussian stretch of the full range of the original image up to a maximum flux level of 99.999\% of all pixel values. Two color JPEG images were created from the HST $V$ and $I$ band optical FITS files using the same software and stretch parameters. The $V$ band image was used for both the blue and green colors, and the $I$ band image was used for the red color. The JPEG images showing the HST footprint on a larger DSS blue optical image were created using Aladin Sky Atlas \citep{Bonnarel2000}. 

All JPEG images that covered the HST field of view were re-sized to be $\sim\textstyle \frac{1}{2}$ page using the software KStudio. The larger GALEX images extending to the 4 times D$_{25}$ distance were resized to be approximately a full page using the same software. Note, therefore, that these larger GALEX JPEG images do not have the same angular scale. 

\section{Summary of the Data Archive \label{summary}}
We have created the panchromatic STARBIRDS archive with reduced observations on a sample of 20 nearby starburst and post-starburst dwarf galaxies. The data are publicly available through the MAST. The suite of data products in the archive include images from GALEX, HST, and $Spitzer$ flux calibrated, background subtracted, registered to the same coordinate system, and cropped to the same fields of view. A second set of the HST images was created by convolving with a 5\arcsec\ PSF and rebinning to a 1.5\arcsec\ pixel scale, equivalent to the GALEX and MIPS 24$\micron$ pixel scale. A third set of GALEX and $Spitzer$ images was created with foreground contamination and background galaxies masked. Finally, a set of larger GALEX images extending to $4\times$D$_{25}$ are also available. Future planned additions to the archive include x-ray, near-IR, H$\alpha$, and \HI\ observations.

Previous results on the processed data in the archive includes in-depth studies of the optical color-magnitude diagrams, resolved star formation histories, measurements of the starburst characteristics, optical characteristics of helium burning stars, and molecular gas content. A multi-wavelength comparison of the star formation properties is presented in a companion paper \citep{McQuinn2015}.

\section{Acknowledgments}
Support for this work was provided by NASA through an ADAP grant (NNX10AD57G), a NASA GALEX grant (NNX10AH98G), and by an NSF Research Experience for Undergraduates (REU) grant (PHYS~0851820) at the University of Minnesota. E.~D.~S. is grateful for partial support from the University of Minnesota. We gratefully acknowledge Steven Warren for helpful advice on image convolution and regridding. The authors would also like to thank the anonymous referee for providing helpful comments that improved this paper. GALEX (Galaxy Evolution Explorer) is a NASA Small Explorer, launched in 2003 April. We gratefully acknowledge NASA's support for construction, operation, and science analysis for the GALEX mission, developed in cooperation with the Centre National d'Etudes Spatiales of France and the Korean Ministry of Science and Technology. This research made use of NASA's Astrophysical Data System, the NASA/IPAC Extragalactic Database which is operated by the Jet Propulsion Laboratory, California Institute of Technology, under contract with the National Aeronautics and Space Administration, and Montage, funded by the NASA's Earth Science Technology Office, Computation Technologies Project, under Cooperative Agreement Number NCC5-626 between NASA and the California Institute of Technology. Montage is maintained by the NASA/IPAC Infrared Science Archive. 

{\it Facilities:} \facility{Hubble Space Telescope; Galaxy Evolution Explorer; \textit{Spitzer Space Telescope}}

\clearpage
\begin{turnpage}
%
\begin{table}
\begin{center}
\caption{Galaxy Sample and Properties}
\label{tab:galaxies}
\end{center}
\begin{center}
\begin{tabular}{llrrrccccrc}
\hline 
\hline 
					&
Alternative			&
R.A.				&
Decl.				&
M$_B$			&
Distance			&
D$_{25}$			&
A$_R$			&
Oxygen			&
					&
Area in			\\
Galaxy			&
Name			&
(J2000)			&
(J2000)			&
(mag)			&
(Mpc)				&
(arcmin)			&
(mag)			&
Abundance		&
Ref.				&
HST FOV			\\
(1)				&
(2)				&
(3)				&
(4)				&
(5)				&
(6)				&
(7)				&
(8)				&
(9)				&
(10)				&
(11)				\\
\hline
\\
\multicolumn{11}{c}{\textbf{Starburst Galaxies}} \\
\\
UGC 4483	&			& 08:37:03.0s 	& $+$69:46:31s 	& $-12.73$   & 3.2  & 1.2  & 0.091    & 7.50 & 1  & 100\%  \\
UGC 6456	& VII~Zw~403	& 11:27:59.9s 	& $+$78:59:39s	& $-14.03$   & 4.3  & 1.5  & 0.096    & 7.64 & 2  & 100\%   \\
DDO 165	 	& 			& 13:06:24.85s	& $+$67:42:25s	& $-15.09$   & 4.6  & 3.5  & 0.065    & 7.80 & 3  & 100\%     \\
IC 4662		& 			& 17:47:08.8s 	& $-$64:38:30s		& $-15.13$   & 2.4  & 2.8  & 0.188    & 8.17 & 4  & 100\%   \\
NGC 6822	& 			& 19:44:56.6s 	& $-$14:47:21s		& $-15.22$   & 0.5  & 15.5 & 0.632   & 8.11 & 5  & 10\%    \\
\\
NGC 4068	& 			& 12:04:00.8s 	& $+$52:35:18s	& $-15.48$   & 4.3  & 3.2  & 0.058    & 7.84 & 3  & 100\%    \\
NGC 2366	& 			& 07:28:54.6s 	& $+$69:12:57s	& $-16.02$   & 3.2  & 7.3  & 0.097    & 8.19 & 6  & 96\%   \\
ESO 154-023	& 			& 02:56:50.38s	& $-$54:34:17s		& $-16.40$   & 5.8  & 8.2  & 0.045    & 8.01 & 3  & 63\%    \\
NGC 784		&			& 02:01:17.0s 	& $+$28:50:15s	& $-16.59$   & 5.2  & 6.6  & 0.158    & 8.05 & 3  & 73\%     \\
Ho II	 		& DDO~50;UGC~4305& 08:19:04.98s& $+$70:43:12s	& $-16.72$   & 3.4  & 7.9  & 0.086    & 7.92 & 7  & 44\%     \\
\\
NGC 4214	& 			& 12:15:39.2s 	& $+$36:19:37s	& $-17.19$   & 2.7  & 8.5  & 0.058    & 8.38 & 8   & 13\%    \\
NGC 5253        & 			& 13:39:55.9s   & $-$31:38:24s		& $-17.38$   & 3.5  & 5.0  & 0.186    & 8.10 & 9   & 100\%    \\
IC 2574         	& 			& 10:28:23.48s  & $+$68:24:43s  	& $-17.46$   & 4.0  &      & 0.096    & 8.07 & 10   & 62\%    \\
NGC 1569	& 			& 04:30:49.0s 	& $+$64:50:53s	& $-18.17$   & 3.4  & 3.6  & 1.871    & 8.19 & 11 & 87\%     \\
NGC 4449	& 			& 12:28:11.9s 	& $+$44:05:40s	& $-18.27$   & 4.2  & 6.2  & 0.051    & 8.32 & 12 & 87\%     \\
\\
\multicolumn{11}{c}{\textbf{Post-Starburst Galaxies}} \\
\\
Antlia Dwarf	& AM~1001$-$270 & 10:04:04.1s& $-$27:19:52s	& $- 9.75$   & 1.3  & 2.0  & 0.212    & 7.39 & 13  & 100\%    \\
UGC 9128	& DDO~187	& 14:15:56.5s 	& $+$23:03:19s	& $-12.71$   & 2.2  & 1.7  & 0.065    & 7.74 & 14 & 100\%     \\
NGC 4163	& 			&12:12:09.1s 	& $+$36:10:09s	& $-14.24$   & 3.0  & 1.9  & 0.052    & 7.69 & 3   & 100\%     \\
NGC 6789	& 			& 19:16:41.1s 	& $+$63:58:24s 	& $-14.32$   & 3.6  & 1.4  & 0.187    & 7.77 & 3   & 100\%    \\
NGC 625		& 			& 01:35:04.6s 	& $-$41:26:10s		& $-16.53$   & 3.9  & 6.4  & 0.044    & 8.10 & 15 & 63\%      \\
\hline
\\
\end{tabular}
\end{center}
\tablecomments{Column (1) and (2) Galaxies are listed according to M$_B$ luminosity, from faintest to brightest. Columns (3) and (4) J2000 coordinates.  Column (5) M$_B$ luminosity corrected for extinction \citep{Karachentsev2004}.Column (6) Distance in Mpc. Column (7) Major axis of M$_B$ 25 mag isophote in arcmin \citep{Karachentsev2004}.  Column (8) A$_R$ (R$=$650 nm) extinction estimates are from the HI maps of \citet{Schlegel1998}.Column (9) and (10) Oxygen abundance and reference. Column (11) Percent of optical disk with the Holmberg radius covered by the $HST$ field of view.}

\tablerefs{(1) \citet{Skillman1994}; (2) \citet{Croxall2009};  (3) L-Z relation; \citet{Lee2006}; (4) \citet{Hidalgo2001a}; (5) \citet{Hidalgo2001b};  (6) \citet{Roy1996}; (7) \citet{Lee2003};  (8) \citet{Kobulnicky1996};  (9) \citet{Kobulnicky1997a};  (10) \citet{Masegosa1991}; (11) \citet{Kobulnicky1997b}; (12) \citet{Skillman1989}; (13) \citet{Piersimoni1999} (14) \citet{vanZee1997}; (15) \citet{Skillman2003};}

\end{table}

\end{turnpage}

\clearpage
\begin{turnpage}
%
\begin{table}
\begin{center}
\caption{Summary of GALEX Observations}
\label{tab:uv_observations}
\end{center}
\begin{center}
\begin{tabular}{lrrrrrrccrr}
\hline 
\hline 
				&
Arch.				&
Arch.				&
				&
Total				&
Total				&
				&
				&
				&
				&
				\\
				&
FUV				&
NUV				&
Archival			&
FUV				&
NUV				&
Final				&
Bkgd. Sub.			&
Masked			&
Flux$\rm{_{FUV}} \times10^{-14}$	&
Flux$\rm{_{NUV}} \times10^{-14}$	\\
Galaxy			&
(ks)				&
(ks)				&
Tile name			&
(ks)				&
(ks)				&
Tile name			&
Method			&
Image				&
(erg s$^{-1}$ cm$^{-2} \AA$) 	&
(erg s$^{-1}$ cm$^{-2} \AA$)	\\
(1)				&
(2)				&
(3)				&
(4)				&
(5)				&
(6)				&			
(7)				&
(8)				&
(9)				&			
(10)				&
(11)				\\
\hline
\\
\multicolumn{11}{c}{\textbf{Starburst Galaxies}} \\
\\
UGC 4483       	& 1.5	 & 1.4	  & NGA$\_$UGC4483	     & 1.5	& 11.2   & GI6\_026202\_UGC4483   &pipeline& Y & 1.7$\pm$0.5 & 0.87$\pm$0.13  \\
UGC 6456       	& 1.6	 & 1.6	  & NGA$\_$VIIZw403	     & 2.1	& 11.6   & GI6\_026203\_UGC6456   &pipeline& Y & 3.0$\pm$0.6 & 1.6$\pm$0.2  \\
      		& 0.4	 & 0.4 	  & AIS$\_117$		     &        	&        &       		  &        &   &      &      \\ 
DDO 165	        & 1.7	 & 1.7	  &  NGA$\_$DDO165	     & 1.7	& 11.3   & GI6\_026208\_DDO165    &pipeline& Y & 6.4$\pm$0.9 & 3.8$\pm$0.3  \\
IC 4662	       	& ...	 & ...	  &       		     & ...	& 9.9    & GI6\_026007\_IC4662    &pipeline& Y & ...	     & 16$\pm$1 \\
NGC 6822       	& 6.3	 & 7.8	  & NGA$\_$NGC6822	     & 6.3	& 10.2   & GI6\_026214\_NGC6822   & custom & Y & 3.7$\pm$0.7 & 1.7$\pm$0.2  \\
      		& 	 &	  &       		     &		&	 & GI6\_026214\_NGC6822   & custom & Y & 4.4$\pm$0.8 & 2.3$\pm$0.2  \\
      		& 	 &	  &       		     &		&	 & GI6\_026214\_NGC6822   & custom & Y & 5.6$\pm$0.9 & 2.3$\pm$0.2  \\
NGC 4068       	& 2.6    & 3.5	  & GI4$\_$095022$\_$NGC4068 & 2.6	& 11.3   & GI6\_026206\_NGC4068   &pipeline& Y & 7.5$\pm$1.0 & 4.1$\pm$0.3  \\
NGC 2366        & 2.9	 & 2.9	  & NGA$\_$NGC2366	     & 2.9	& 7.2    & GI6\_026210\_NGC2366   &pipeline& Y & 31$\pm$2    & 17$\pm$ 1 \\
ESO 154$-$023   & 1.7	 & 1.7	  & GI4$\_$095006$\_$ESO154$\_$G023& 1.7& 9.9    & GI6\_026209\_ESO154m023&pipeline& Y & 9.1$\pm$1.1 & 5.1$\pm$0.3   \\
NGC 784	       	& 2.2 	 & 2.2	  & GI4$\_$095004$\_$NGC0784 & 2.2	& 7.6    & GI6\_026212\_NGC784    &pipeline& Y & 6.8$\pm$0.9 & 4.7$\pm$ 0.3  \\
Ho II	        & 21.1	 & 21.1	  & GI3$\_$050003$\_$HolmbergII&22.8   	& 22.8   & GI6\_026218\_HolmbergII& custom & Y & 32$\pm$2    & 17$\pm$1 \\
      		& 1.7	 & 1.7 	  & NGA$\_$HolmbergII	     &        	&        &       		  &        &   &      &      \\ 
NGC 4214       	& 2.0	 & 2.0	  & GI4$\_$095023$\_$NGC4214& 2.0	& 10.2   & GI6\_026215\_NGC4214   & custom & N & 54$\pm$3    & 33$\pm$1	 \\
NGC 5253	&1.7	 & 1.7	  & GI4$\_$095049$\_$NGC5253 & 2.3	& 4.5    & GI6\_026213\_NGC5253   &pipeline& Y & 33$\pm$2    & 24$\pm$1 \\
      		& 0.6	 & 0.6 	  & NGA$\_$NGC5253	     &        	&        &       		  &        &   &      &      \\ 
IC 2574		& 13.0	 & 13.0	  & GI3$\_$050005$\_$IC2574  & 15.0   	& 16.5   & GI6\_026219\_IC2574    & custom & Y & 34$\pm$2    & 18$\pm$1 \\
      		& 1.9	 & 3.5 	  & NGA$\_$IC2574	     &        	&        &       		  &        &   &      &      \\ 
NGC 1569	& 7.1	 & 7.1	  & NGA$\_$NGC1569	     & 7.1	& 7.1    & NGC\_NGC1569	     	  &pipeline& Y & 3.2$\pm$0.7 & 2.8$\pm$0.2  \\
NGC 4449 	& 3.3	 & 1.7	  & GI4$\_$095051$\_$NGC4460 & 4.2	& 15.5   & GI6\_026216\_NGC4449   & custom & Y & 160$\pm$5   & 85$\pm$1	 \\
      		& 0.9	 & 0.9 	  & NGA$\_$NGC4449	     &        	&        &       		  &        &   &      &      \\ 
\\
\multicolumn{11}{c}{\textbf{Post-Starburst Galaxies}} \\
\\
Antlia Dwarf    & 13.8   & 17.2   & NGA\_Antlia\_Dw	     & 13.8   	& 17.2   & NGA\_Antlia\_Dw	  & custom & Y & 0.33$\pm$0.22 & 0.39$\pm$0.09  \\
UGC 9128       	& 2.6	 & 5.8	  & GI4$\_$016007$\_$DDO187  & 2.6    	& 7.6    & GI6\_026201\_UGC9128   &pipeline& Y & 1.2$\pm$0.4 & 0.82$\pm$0.13  \\
NGC 4163       	& 1.7	 & 1.7	  & GI3$\_$061001$\_$NGC4163 & 12.1   	& 15.6   & GI6\_026217\_NGC4163   &pipeline& Y & 2.5$\pm$0.5 & 1.8$\pm$0.2  \\
      		& 10.3 	 & 13.4	  & GI4$\_$015004$\_$NGC4163 &        	&        &       		  &        &   &      &      \\ 
NGC 6789      	& 1.5	 & 1.5	  & NGA$\_$NGC6789	     & 1.5	& 2.9    & GI6\_026204\_NGC6789   &pipeline& Y & 0.8$\pm$0.3 & 0.64$\pm$0.11 \\
NGC 625	       	& 4.5	 & 4.5	  & GI1$\_$047007$\_$NGC0625 & 4.5	& 11.3   & GI6\_026211\_NGC625    & custom & N & 9.7$\pm$1.2 & 6.2$\pm$0.4  \\
\hline
\\
\end{tabular}
\end{center}
\tablecomments{Column (1) Galaxy name. Columns (2) and (3) FUV and NUV integration time for the GALEX archival observations. Column (4) Tile name for archival observations. Columns (5) and (6)Total FUV and NUV integration time for the GALEX observations including the archival observations plus new legacy observations (PID 60026). Column (7) Final tile name for the GALEX co-added observations. Column (8) Background subtraction was performed as part of the GALEX v7.1 pipeline for all but 6 galaxies; the remaining 6 galaxies (including 3 fields of view for NGC~6822) showed emission structure from the galaxy or cirrus contamination in the sky background images and thus the background subtracted intensity images were over- or under-subtracted. We measured and subtracted the background separately for these systems as described fully in Section 4.1. Column (9) Foreground and background contamination were masked from the GALEX images as noted. Columns (10) and (11) Integrated FUV and NUV flux measurements from the GALEX images in the HST field of view with appropriate foreground and background contamination masked. Uncertainties assume a poisson distribution and are based on the square root of the number of counts at each waveband.}

\end{table}

\clearpage
\end{turnpage}

%
\begin{table}
\begin{center}
\caption{Summary of $HST$ Observations}
\label{tab:hst_observations}
\end{center}
\begin{center}
\begin{tabular}{lrrrrrrccrr}
\hline 
\hline 
				&
HST				&
				&
HST				&
No. of			&
F555W				&
F606W				&
F814W				\\
Galaxy			&
Proposal ID			&
PI				&
Instrument			&
Pointings			&
(s)				&
(s)				&
(s)				\\
(1)				&
(2)				&
(3)				&
(4)				&
(5)				&
(6)				&
(7)				&
(8)				\\
\hline
\\
\multicolumn{8}{c}{\textbf{Starburst Galaxies}}\\
\\
UGC 4483       & 8769 	& Thuan  		& WFPC2	& 1  	&9500 	& ...   & 6900		 	\\
UGC 6456       & 6276	& Westphal  	& WFPC2	& 1  	&4200 	& ...   & 4200			\\
DDO 165	       & 10605	& Skillmnan 	& ACS	& 1   	&4800 	& ...   & 4800			\\
IC 4662	       & 9771 	& Karachentsev  & ACS	& 1  	& ... 	& 1200  & 900 				\\
NGC 6822       & 6813 	& Hodge		& WFPC2	& 3   	&3900 & ...   & 3900				\\
\\
NGC 4068       & 9771 	& Karachentsev  & ACS	& 1   	& ... 	& 1200  & 900 				\\
NGC 2366       & 10605	& Skillman  	& ACS	& 2   	&4800 	& ...   & 4800			\\
ESO 154-023    & 10210	& Tully  	& ACS	& 1   	& ... 	& 1000  & 1300				      	\\
NGC 784	       & 10210	& Tully  	& ACS	& 1   	& ... 	& 930   & 1200				      	\\
Ho II	       & 10605	& Skillman	& ACS	& 2   	&4700 	& ...   & 4700				\\
\\
NGC 4214       & 6569 	& MacKenty  	& WFPC2	& 1   	&1300 	& ...   & 1300			\\
NGC 5253       & 10765	& Zezas	  	& ACS	& 2   	&2400 	& ...   & 2400	  		\\
IC 2574	       & 9755 	& Walter  	& ACS	& 3   	&4800 & ...  & 4800 					\\
NGC 1569       & 10885	& Aloisi  	& ACS	& 1   	& ... 	& 61400 & 23700				\\
NGC 4449       & 10585	& Aloisi  	& ACS	& 2   	&4900 	& ...   & 4100				\\
\\
\multicolumn{8}{c}{\textbf{Post-Starburst Galaxies}} \\
\\
Antlia Dwarf   & 10210	& Tully  	& ACS	& 1  	& ... 	& 990   & 1170				      	\\
UGC 9128       & 10210	& Tully  	& ACS	& 1  	& ... 	& 990   & 1170				      	\\
NGC 4163       & 9771 	& Karachentsev 	& ACS	& 1  	& ... 	& 1200  & 900 			\\
NGC 6789       & 8122 	&Schulte-Ladbeck& WFPC2	& 1  	&8200 	& ...   & 8200		\\
NGC 625	       & 8708 	& Skillman  	& WFPC2	& 1   	&5200 	& ...   & 10400			\\
\hline
\\
\end{tabular}
\end{center}

\tablecomments{The observations of NGC~6822 include 3 non-overlapping pointings. Exposure time for one of three, non-overlapping fields of view is greater, totaling 6400 secs for each filter.}

\end{table}

%
\begin{table}
\begin{center}
\caption{Summary of $Spitzer$ MIPS Observations}
\label{tab:mips_observations}
\end{center}
\begin{center}
\begin{tabular}{lllrrrrrrccc}
\hline 
\hline 
				&
$Spitzer$			&
				&
$24 \micron$			&
$70 \micron$			&
$160 \micron$			&
Flux$_{24 \micron}$		&
Flux$_{70 \micron}$		&
Flux$_{160 \micron}$		&
24$\micron$			&
70$\micron$			&
160$\micron$			\\
Galaxy			&
Program ID			&
PI				&
(s)				&
(s)				&
(s)				&
(Jy)				&
(Jy)				&
(Jy)				&
Masked			&
Masked			&
Masked			\\
(1)				&
(2)				&
(3)				&
(4)				&
(5)				&
(6)				&
(7)				&
(8)				&
(9)				&
(10)				&
(11)				&
(12)				\\
\hline
\\
\multicolumn{12}{c}{\textbf{Starburst Galaxies}} \\
\\
UGC 4483       & LVL 	& Kennicutt 	& 160	& 80	&  16   & 0.07 	& 0.1 	& ND 	& Y &  N & N \\
UGC 6456       & 59	& Rieke	  	& 3	& ...	&  10   & 0.03	& ...	& ND	& Y &  ...& N \\
DDO 165	       & LVL 	& Kennicutt 	& 160	& 80	&  16   & ND	& 0.06	& ND 	& Y &  Y& N \\
IC 4662	       & 3567	& Vacca		& 3	& 10	& ...   & $>$5.0& $>9.0$& ...	& Y &  N & ... \\
NGC 6822       & SINGS 	& Kennicutt 	& 160	& 80	&  16   & 0.2	& 3.2	& 6.0	& Y &  N & N \\
	       &	&		&	&	&	& 0.01	& 3.2	& 6.9	& N  &  N & N \\
	       &	&		&	&	&	& 0.2	& 3.2	& 4.0	& N  &  N & N \\
\\
NGC 4068       & LVL 	& Kennicutt 	& 160	& 80	&  16   & 0.02	& 0.7	& 1.0 	& Y &  N & N \\
NGC 2366       & LVL 	& Kennicutt 	& 160	& 80	&  16   & 0.7	& 5.0	& 3.7	& Y &  Y& Y \\
ESO 154-023    & LVL 	& Kennicutt 	& 160	& 80	&  16   & 46	& 0.7	& 1.1	&  N &  N & N \\
NGC 784	       & LVL 	& Kennicutt 	& 160	& 80	&  16   & 0.03	& 1.1	& 1.4	& Y &  N & N \\
Ho II	       & LVL 	& Kennicutt 	& 160	& 80	&  16   & 0.2	& 2.7	& 3.0	& Y &  N & N \\
\\
NGC 4214       & LVL 	& Kennicutt 	& 160	& 80	&  16   & 1.7	& 17	& 19	& N  &  N & N \\
NGC 5253       & LVL 	& Kennicutt 	& 160	& 80	&  16   & 8.8	& 26	& 20.	& Y &  N & N \\
IC 2574	       & LVL 	& Kennicutt 	& 160	& 80	&  16   & 0.2	& 3.6	& 6.9	& Y &  N & N \\
NGC 1569       & 69	& Fazio 	& 3 	& ...	&  3    & 7.4	& ... 	& ...	& N  &  ...& N \\
NGC 4449       & LVL 	& Kennicutt 	& 160	& 80	&  16   & 3.1	& 41	& 71	& Y &  N & N \\
\\
\multicolumn{12}{c}{\textbf{Post-Starburst Galaxies}} \\
\\
Antlia Dwarf   & LVL 	& Skillman 	& 107	& 42	&  8.4  & ND 	& ND 	& ND 	& N  &  N & N \\
UGC 9128       & LVL 	& Kennicutt 	& 160	& 80	&  16   & ND 	& ND 	& ND 	& Y &  N & N \\
NGC 4163       & LVL 	& Kennicutt 	& 160	& 80	&  16   & 0.003	& 0.07 	& 0.2 	& Y &  N & N \\
NGC 6789       & 40301	& Fazio		& 10	& ...	& ...   & 0.008	& ... 	& ...	& Y &  ...& ... \\
NGC 625	       & LVL 	& Kennicutt 	& 160	& 80	&  16   & 0.8	& 5.6	& 5.8	&  N &  N & N\\
\hline
\\
\end{tabular}
\end{center}
\tablecomments{Column (1) Galaxy name. Columns (2) and (3) Spitzer program ID and PI. Columns (4)-(6) Exposure times for the LVL and SINGS data are the approximate times for the central pixel in each map. Columns (7)-(9) Flux measurements in the HST field of view after any evident foreground and background contamination has been masked. ND is a non-detection at that wavelength. Lower limits on the flux are noted for IC~4662 where the MIPS observations did not cover the HST field of view. Columns (10)-(12) Foreground and background contamination were identified and appropriately masked from the MIPS images as noted.}

\end{table}

\end{document}